\documentclass{IEEEcsmag}

\usepackage[dvipsnames]{xcolor}

\usepackage[colorlinks,urlcolor=blue,linkcolor=blue,citecolor=blue]{hyperref}

\usepackage{upmath}
\usepackage{comment}
\usepackage{tabularx}
\usepackage{soul}
\usepackage{xcolor}

\sethlcolor{Goldenrod}

\jvol{XX}
\jnum{XX}
\paper{8}
\jmonth{May/June}
\jname{IT Professional}
\pubyear{2021}

\begin{document}

\sptitle{DEPARTMENT: Viewpoints}

\title{Healthy skepticism in AI: a data visualization research agenda}

\author{G.\ Elisabeta Marai}
\affil{University of Illinois Chicago, Chicago, IL, 60607, U.S.A.}

\author{Marc Baaden}
\affil{CNRS (French National Centre for Scientific Research), Paris, 75794, France}

\author{Michael Behrisch}
\affil{Utrecht University, Utrecht, 3584CC, The Netherlands}

\author{Michael Krone}
\affil{Stuttgart Technical University of Applied Sciences, Stuttgart, 70174, Germany}

\author{Pere-Pau Vázquez}
\affil{Universitat Politècnica de Catalunya, Barcelona 08034, Spain}

\markboth{VIEWPOINTS}{VIEWPOINTS}

\begin{abstract}
 Research in data visualization of artificial intelligence (AI) models
 has historically focused on enhancing trust through visual explanations of AI. The trustworthiness line of work was built at least partially on an assumption that humans were critical users {\textcolor{Black}{unlikely to adopt AI technology}}. It is increasingly clear that human trust levels in AI span, in fact, a wide range from critical to over-reliant. There is an urgent need to support both trust and healthy skepticism in AI solutions. We argue that it is healthy for humans to adopt a skeptical view both on the results of AI models and on the use of such AI models. We share our thoughts on {\textcolor{Black}{the rising phenomenon of over-reliance on AI models, the risks and opportunities in using AI models,}} and the role of data visualization in over-reliance situations where humans are not motivated to engage in critical thinking.

\end{abstract}

\maketitle

%
%

\chapteri{R}{\textcolor{Black}{esearch in data visualization of artificial intelligence (AI) models has traditionally centered on visual explanations of AI, with the explicit mission "to enhance trust in AI models"~\cite{van2023flow, chatzimparmpas2020state, chatzimparmpas2024visualization}.}} The trustworthiness line of work was partially based on the assumption and early evidence, some from the biomedical field~\cite{vanlent2004explainable}, that humans were critical users {\textcolor{Black}{unlikely to adopt AI technology}}. 
{\textcolor{Black}{Further research in algorithm aversion suggested that user trust could be increased by making models more transparent and giving users greater control, for example by allowing minor adjustments to model outputs or customization of model inputs (see the first sidebar for recommended readings).}}
{\textcolor{Black}{This viewpoint article focuses on how visualization can help calibrate trust in AI, extending established work on visualization for trust enhancement~\cite{chatzimparmpas2020state, chatzimparmpas2024visualization, van2023flow} to address over-reliance. We recognize that AI also has broader applications within visualization, beyond explainable AI (XAI), but these are outside the scope of this article.}}


\textcolor{Black}{This work is an outcome of discussions at Dagstuhl Seminar 26101, a venue for in-depth, collaborative research. Bringing together researchers in this setting
enabled sustained and detailed exploration of the challenges surrounding data visualization for AI explainability (VXAI). Through these discussions, we identified the topic of over-reliance on AI models as a significant emerging issue that warrants broader attention (Fig.~\ref{fig:vxai}).} Throughout this article, we use the term AI models broadly, encompassing not only generative AI (genAI) but also \textcolor{Black}{non-genAI} models for prediction, analysis, and other functional tasks. These non-genAI models focus on pattern recognition, data analysis, and decision-making rather than creating new content.

{\textcolor{Black}{Drawing on our collective experience and prior work on building new AI models for practical applications~\cite{luciani2020spatial, wentzel2019cohort, wentzel2024ditto, nipu2025lessons}, these discussions highlighted an urgent need to support both trust and
healthy skepticism in AI solutions. Critically, we identified the data visualization community as particularly well positioned, given its knowledge, experience, and tools, to support AI models as more effective collaborators for humans. The ideas emerging from this working group were further refined through discussions with colleagues and collaborators. }

In this article, we share our thoughts on {\textcolor{Black}{the rising phenomenon of over-reliance on AI models, and the risks and opportunities in using AI models}}. This evaluation allows us to assess when to deploy AI appropriately rather than rejecting useful applications through blanket skepticism or causing harm through uncritical adoption. We then articulate reasons to practice healthy skepticism in AI models. Last, we present a research agenda for data visualization in this context.

\begin{figure*}
\centerline{\includegraphics[width=35pc]{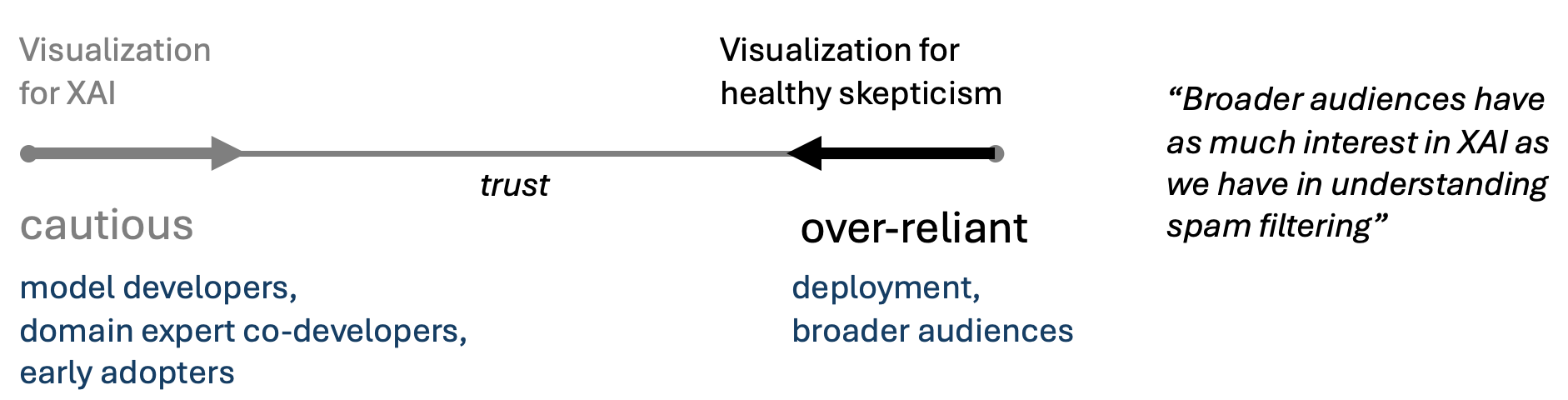}}
\caption{Visualization for healthy skepticism and visualization for explainable AI (XAI). The space of reliance and trust is a spectrum, from cautious or avoidant (left) to over-reliant, nearly blind trust (right). Visualization for XAI is focused on the left end---its mission is to increase trust, and it tends to serve model builders, co-builders, or early adopters, who have an interest in the inner workings of the AI model and are motivated to interact with it and explore. Visualization for healthy skepticism aims to move in the opposite direction, from the opposite end, and needs to address a broader, usually deployment (as opposed to development) audience. Most broader audiences have as much interest in XAI as we, the authors, have in understanding spam filtering. Visualization can help in over-reliance situations where humans are not motivated to engage in critical thinking.}
\label{fig:vxai}
\end{figure*}

\section{Emerging AI over-reliance}
\textcolor{Black}{Users can over-rely on automated AI systems, often ignoring contradictory evidence or failing to verify outputs. This is a major, growing risk which can result in security, legal (e.g., legal system fabrications), or safety (e.g., medical misdiagnoses) failures.  To understand the context of over-reliance, we draw on examples from our own work.}

\noindent\textcolor{Black}{
\textbf{Instance 1.} For the past decade, the first author (GEM) has led an interdisciplinary team of oncologists and data scientists. Together, we have built a powerful, state-of-the-art set of AI-enabled systems to predict survival and treatment side-effects in head and neck cancer.  Some of these non-genAI tools have been validated  internationally using data from multiple major medical centers. The team is increasingly aware that clinicians around the world, as well as patients and their families, are consulting the resulting publicly available, interactive visualization systems for decision support. The team's clinicians attribute this trust to the models' validation and demonstrated performance, noting that they ``have been rigorously validated on large datasets from multiple international centers and have demonstrated good performance.'' This explanation echoes van den Elzen et al.'s~\cite{van2023flow} suggestion that users may place ``faith in the authority or organization behind these models.''}


\noindent\textcolor{Black}{{\textbf{Instance 2}}. In 2024, GEM's visual computing group started using an embedded deep learning-based cell segmentation model (\emph{bin2cell}) to  meet specific challenges in biological data visualization~\cite{zhao2026loom}. In this case, we (GEM's group) have not checked the credentials or validation of \emph{bin2cell}. Arguably, we trust \emph{bin2cell} because it provides a functionality we need, and it fulfills this functionality reasonably well. }

\noindent\textcolor{Black}{{\textbf{Instance 3}}. In 2025, GEM's group created a new attention-based, self-supervised neural network architecture for detecting regions of interest in 3D microscopy images. There was eager interest from multiple organizations in using the resulting non-genAI visual analysis solution, months before its validation was completed and published~\cite{fathollahian2026congat}. Arguably, people want to use this AI model because it provides a desired functionality, and works well enough.}

\noindent\textcolor{Black}{{\textbf{Instance 4 and more}}. In 2025, GEM's group noted clinicians trusting unconditionally predictions from a non-genAI digital twin for healthcare, powered by a visual analysis interface~\cite{wentzel2024ditto}, in cases where our team knew some of those predictions were incorrect.} The same year, PPV's group observed physicians trusting a colon segmentation tool tested on three imaging machines and a relatively large dataset, while being likely unaware of the potential limitations in terms of sample variety~\cite{ORELLANA2025102528}.

In 2026, beyond biomed data visualization, a German court ruled that Google is directly liable for false statements produced by its AI Overviews feature. The court dismissed the argument that users knew anyway that "AI-generated information should not be trusted blindly" (see first sidebar readings). 

The same year, AAAI, the Association for the Advancement of Artificial Intelligence, officially deployed a genAI-assisted peer review pilot program to enhance the academic paper review process for the AAAI-26 conference, even though all genAI models have a significant risk of hallucinating when they generate a summary. 
These last two cases indicate that beyond experts, broader audiences also experience over-reliance on AI models.} 

\subsection{AI range and a growing gap}
\textcolor{Black}{The above instances indicate that data visualization uses many valuable non-genAI models, which provide valuable support for pattern recognition, data analysis, and decision-making. Furthermore, these instances illustrate emerging over-reliance on both non-genAI and genAI models, potentially fueled by the surging popularity of generative AI.
}

\textcolor{Black} {We also witness a growing gap between what computer science researchers know about AI models, and how the public perceives AI models. Some of our relatives believe, for example, that an AI model could refuse to make a wrong prediction. In reality, AI models do not know inherently when they are wrong. Some of our students do not understand why clinician de-skilling is a problem. In reality, AI models experience performance drift. Some of our friends believe that the next generation of genAI models will overcome all current limitations, or that visualization for explainable AI may stop being relevant. In reality, there are good reasons to be skeptical even of a perfectly functioning AI model, e.g., data training limitations (see the Skepticism section below).}

Current visualization research in the XAI area assumes the human is motivated to work in a loop with an AI model. We argue that that is no longer correct. This is not incremental data visualization work: the topic we introduce is largely unexplored and underserved. Visualization can not only facilitate understanding of AI algorithms by a motivated human-in-the-loop. It can also help calibrate trust, enable the detection of mistakes by AI algorithms, and enhance decision making processes in over-reliance situations where humans are not motivated to engage in critical thinking.

\textcolor{Black}{Beyond delegation to a higher authority, people trust AI for a complex mix of psychological, social, and practical reasons. These reasons range from incremental normalization, i.e., gradual integration of AI into daily life through recommendations, autocomplete, or search results, to difficulty in detecting errors, where plausible-sounding but incorrect information is hard to identify without domain expertise or fact-checking, and to perceived non-judgment from the AI, particularly in contexts involving vulnerability, stigma, or social evaluation. Arguably, people use these AI models first and foremost because they provide desired functionality, and they work reasonably well.
}

\section{AI models supporting activities}

There is support for the activity-support basis of AI model acceptance. Both HCI research in activity-centered-design~\cite{norman2013design} and data visualization research~\cite{marai2017activity} show that humans quickly adopt those tools that provide desired activity support. Some of the most important ways in which AI models enhance human skills while supporting activities, and can thus be extremely attractive, can be seen in our own data visualization field, where AI models can, \textcolor{Black}{for example}:
\begin{itemize}
    \item \textbf{Extend cognitive reach}, by serving as external memory and computational support \textcolor{Black}{(e.g., non-genAI Instance 1 above).}
    \item \textbf{Accelerate routine tasks}, by handling repetitive data processing or formatting \textcolor{Black}{ (e.g., non-genAI Instance 2 above).}
    \item \textbf{Augment pattern recognition at scale}, by processing vast datasets to identify trends humans could miss \textcolor{Black}{(e.g., non-genAI Instance 3 above).}
    \item \textbf{Enable rapid prototyping}, by accelerating the creation of drafts, code, designs, and models \textcolor{Black}{(e.g., genAI in Zhang et al.~\cite{zhang2025auragenome}).}
    \item \textbf{Enhance brainstorming and ideation}, by generating diverse perspectives and alternatives quickly, and helping humans overcome mental blocks \textcolor{Black}{(our students use genAI for this purpose).}
    \item \textbf{Democratize expert knowledge access}, by making specialized information accessible to non-experts and lowering barriers to interdisciplinary work \textcolor{Black}{(we, the authors, sometimes use genAI for this purpose).}
    \item \textbf{Support personalized learning paths}, by adapting to individual learning styles, pacing, and knowledge gaps \textcolor{Black}{(some of our teaching-track colleagues use genAI successfully for this purpose).}
\end{itemize}

There are clearly appropriate AI uses that complement human skills without impairing development of critical skills, e.g.: being a virtual brainstorming partner, enhancing human capabilities, checking work, or exploring concepts after attempting to solve problems.  

\section{AI models damaging human skills}

At this point, however, there is also mounting evidence that AI models can damage human skills (see second sidebar recommended readings). We witness this phenomenon as researchers and as educators, and note  (\textcolor{Black}{the list below is not exhaustive)} that AI models can, \textcolor{Black}{for example}:
\begin{itemize}
    \item \textbf{Atrophy critical thinking and mathematical intuition}, where both domain experts and students accept AI outputs without verification or independent reasoning, or fail to develop pattern recognition {\textcolor{Black}{ (we see this phenomenon in the over-reliance Instance 4.)}}
    \item \textbf{Reduce memory and recall}, as humans no longer engage in the encoding processes that build durable knowledge {\textcolor{Black}{(we see this phenomenon in our reduced ability to spell.)}}
    \item \textbf{Atrophy pattern recognition}, as students fail to develop pattern recognition, and domain experts lose skills critical to their job; this phenomenon is now known as {\emph {de-skilling}} (see second sidebar readings). 
    \item \textbf{Erode writing proficiency, reading comprehension, and information literacy}, where delegating these tasks to AI reduces practice in organizing thoughts, evaluating sources, and distinguishing credible information; in turn, this prevents the development of interpretative abilities and weakens fundamental communication skills.
    \item \textbf{Weaken collaborative skills}, where using AI as a substitute for peer interaction reduces practice in negotiation, perspective taking, constructive disagreement, and the social skills essential for teamwork.
    \item \textbf{Diminish problem-solving abilities}, where students miss developing persistence, creative approaches, and the cognitive skills gained from working through challenges. 
    \item \textbf{Reduce efficiency}, by increasing the amount of effort to verify the AI output (see second sidebar readings).
    \item \textbf{Impair creativity and originality}, by reducing exposure to the cognitive discomfort that sparks genuine innovation and novel connections (anecdotally, we increasingly see this in classrooms).
\end{itemize}

There are clearly skill-damaging AI uses: generating first drafts without planning, solving problem sets without attempting a solution, replacing reading with summaries, or usage that leads to atrophy of other critical skills. {\textcolor{Black}{Looking at the second list versus the first, AI tools may erode the very skills they assist.}}

More generally, and particularly in high-risk situations, there is inherent value in supporting slow thinking~\cite{kahneman2011thinking} (see first sidebar recommended readings). {\textcolor{Black}{In his book "Thinking, fast and slow", Nobel laureate Kahneman described two distinct systems that drive human decisions: thinking system~1 is fast, intuitive, and emotional, while thinking system~2 is slower, more deliberative, and logical. Kahneman's book demonstrates that our minds are highly prone to biases and cognitive errors because thinking system~1 leads most of the time.} Appropriate AI reliance could help humans avoid fast, intuitive, emotional decisions (i.e., thinking system~1) in favor of slow, deliberate, and logical thinking (i.e., thinking system~2). The use of AI at different stages can either support or undermine learning objectives, skill development, and slow thinking. }

\section*{Recommended Readings (Sidebar 1)}
\hl{\noindent\textbf{Thinking fast and slow:} Kahneman (2011) outlines two distinct systems that drive our decisions: system~1 is fast, intuitive, and emotional, while system~2 is slower, more deliberative, and logical. The book demonstrates that our minds are highly prone to biases and cognitive errors because system~1 often leads:\\
R1. Kahneman, D. ``Thinking, fast and slow,'' (2011).}\vspace{.5em}

\noindent\hl{\textbf{Google liable for fake AI answers: }
In a 2026 lawsuit against the tech company's use of genAI overviews, a German court rejected the defense line that users knew anyway that ``AI-generated information should not be trusted blindly.'' The reader was given no indication of any unreliability in the content, the court found:\\
R2. Connor, R., ``German court holds Google liable for fake AI answers.'' (2026).}\vspace{.5em}

\noindent\hl{\textbf{Activity centered design: }
In the third edition (2013)  of ``The Design of Everyday Things,'' Norman departs from the human-centered design paradigm he described in the first edition, and discusses activity-centered design as a stronger predictor of client tool adoption:\\
R3. Norman, D., ``The design of everyday things: Revised and expanded edition,'' (2013).}\vspace{.5em}

\noindent\hl{\textbf{Algorithm aversion: }
Algorithm aversion (2015) is the psychological tendency for humans to distrust or reject advice from an algorithm or AI, even when it is proven to outperform human judgment:\\
R4. Dietvorst, B. et al., ``Algorithm aversion: people erroneously avoid algorithms after seeing them err.'' (2015).} \vspace{.5em}

\noindent\hl{\textbf{Overcoming algorithm aversion: }
Overcoming algorithm aversion (2016) is the process of reducing human reluctance to use data-driven algorithms, often by giving users control to slightly modify or select the algorithms they use:\\
R5. Dietvorst, B. et al., ``Overcoming algorithm aversion: People will use imperfect algorithms if they can (even slightly) modify them.'' (2018).}\vspace{.5em}

\section{Healthy skepticism in AI models}

Even though both non-genAI and genAI models, when used at appropriate stages, can clearly augment human cognitive skills, healthy skepticism remains warranted: even useful models may produce, for instance, incomplete results. 
Most AI solutions can also inherit data-biases and human-biases, and can produce incorrect or misleading information with high confidence. For example, the IBM Watson for Oncology AI solution was designed to analyze vast amounts of medical data to suggest personalized cancer treatments, but it ultimately failed due to inaccurate, unsafe, and biased recommendations (see second sidebar). Critical decision-making and accountability are especially required in domains such as healthcare, law, policymaking, engineering, construction, or architecture. A non-exhaustive list of the reasons to entertain healthy skepticism in AI models includes:
\begin{itemize}
    \item \textbf{Generalization failures}, as AI models only learn from historical data, where they inherit biases, gaps, and errors, and have blind spots in novel or underrepresented scenarios (e.g., over-reliance Instance 4). 
    \item \textbf{Embedded social biases}, as AI without explicit specialized programming can amplify societal prejudices present in training data, and are known to perpetuate discrimination in hiring, lending, criminal justice, and healthcare. 
    \item \textbf{Adversarial vulnerability}, where small, intentional input perturbations often invisible or imperceptible to humans can cause dramatic  exploitability in security-critical applications.
    \item \textbf{Hallucination and confabulation}, as genAI (generative) models confidently produce, by their probabilistic construction approach, plausible-sounding information that nevertheless can be factually incorrect.
    \item \textbf{Optimization misalignment}, where AI models optimize for and achieve high technical metrics such as area under the curve, that may not align with human values or intended outcomes (e.g., accuracy in the clinic), missing the actual goal or causing unintended harm.
    \item \textbf{Reduced causal understanding}, as many AI models identify correlations, not causation, and models may exploit spurious patterns that work in training but fail when underlying relationships change.
    \item \textbf{Temporal degradation}, as model performance decays with real-world data distribution evolution; concept drift means yesterday’s accurate model becomes today’s unreliable predictor unless continuously monitored.
    \item \textbf{Accountability limitations}, as deep learning models are “black boxes”, making it difficult to verify reasoning, debug errors, or ensure accountability; this issue is particularly poignant in the life sciences, domains that have strong restrictions in terms of ethical considerations and auditability requirements, where, for example, the doctor is responsible for the health of the patient by law and can hence not freely offload responsibility to AI.
    \item \textbf{Overconfidence}, as AI models seldom reliably assess their own limitations, uncertainty, or knowledge boundaries.
\end{itemize}

\section{VXAI and AI over-reliance}
\textcolor{Black}{How can data visualization mitigate blind trust in AI models? It might be helpful to consider first the data visualization background. First, there are two main types of data visualization: {\textbf{explanatory visualization}}, which are data communication techniques that present specific, curated insights to an audience to tell a clear, concise story. These techniques are designed to guide the viewer towards a specific conclusion, for example through the design of charts. The second type is {\textbf{exploratory visualization}}, which is an open-ended, interactive, iterative process to examine data visually to discover patterns, trends, anomalies, or relationships.}

\textcolor{Black}{Second, within data visualization, there has been remarkable work in the area of AI explainability (XAI). The visualization for XAI (VXAI) area is active, as captured in two surveys in 2020 and 2024~\cite{chatzimparmpas2020state, chatzimparmpas2024visualization}. To briefly summarize this area, VXAI enhances trust via increased model transparency and increased control. These type of enhancements typically require interaction. Thus, from the two types of visualization, VXAI work falls typically squarely in the second type, exploratory visualization.
 }

\textcolor{Black}{We can think of the space of reliance and trust as a spectrum, from cautious or avoidant (left) to over-reliant, nearly blind trust (right) in Fig.~\ref{fig:vxai}. VXAI is focused on the left end---its explicitly stated mission is to increase trust~\cite{chatzimparmpas2020state}. This makes sense, because much of the VXAI work comes from data visualization groups working with model developers, domain expert co-developers, or early adopters. This class of clients have an interest in the inner workings of the AI model and are motivated to interact with it and explore.}
 
\textcolor{Black}{To support healthy skepticism, we must address the opposite end of the spectrum, reducing over-reliance without pushing users to reject AI altogether. Healthy skepticism needs to address a broader, usually deployment (as opposed to development) audience. Examples of broader audiences may be oncologists, patients, and even the public at large. One's own experience may vary, but our group's experience has been that {\textbf{most broader audiences have as much interest in XAI as we, the authors, have in understanding spam filtering}}. To clarify, spam filtering is nowadays conducted by AI models. It works well 90\% of the time, and we can handle its failures. If the failures get too annoying, we typically move to a different platform. We are not motivated to interact with or dig into the spam filtering model, not even when we see it fail. This echoes our experience with over-reliance Instance 4. For example, the digital twin instance provides an XAI gallery of most similar patients. This gallery captures well the model mechanics, which identify and leverage subcohorts of spatially-similar patients. This VXAI feature was a favorite of co-builders, yet broader audiences neither used it nor saw its value.}

\textcolor{Black}{In short, if as a research community, we expect a user who exhibits nearly blind trust to use interactive VXAI with no explicit prompting, we have lost the over-reliance battle already. We could also replace “healthy skepticism” with “critical reflection”, “informed scrutiny”, or “appropriate caution”, but these terms do not capture as well our respect for non-genAI models and our caution towards AI. As shown in the sections above, AI can be very helpful when used appropriately.}

\begin{figure}
\centerline{\includegraphics[width=17pc]{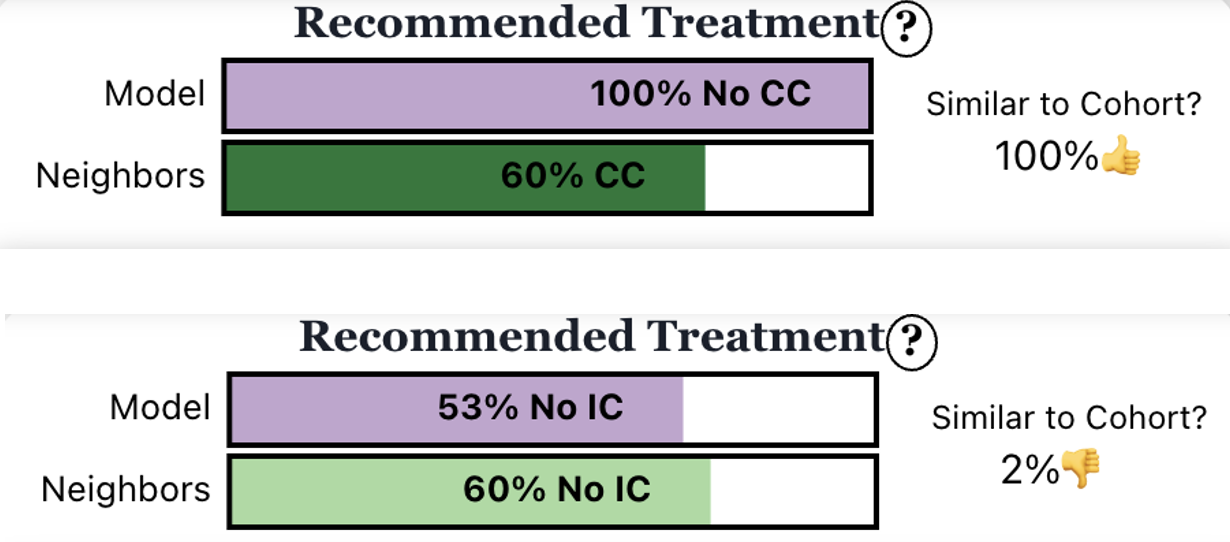}}
\caption{Explanatory design for healthy skepticism from Wentzel et al.~\cite{wentzel2024ditto}. A yellow thumbs up/down icon indicates the AI model limitations: how much data support the digital twin model had when making that recommendation. Along with the digital twin AI model prediction in purple, a decoy AI model prediction based on nearest neighbors is shown in green, to further encourage the clinician to pause and reflect. }
\label{fig:wentzel}
\end{figure}

\section{A research agenda}
Through its long tradition of interdisciplinary collaboration, the data visualization community is uniquely equipped with the necessary knowledge, experience, and tools to help \textcolor{Black}{combat over-reliance on AI models. Here is a first agenda for healthy skepticism research in visualization, where the purpose of an agenda is to define research directions and pose research questions but not yet give answers to these questions:}

\subsection{Model transparency, without interaction} \textcolor{Black}{ We need new encodings and new designs to help users question, challenge, or
appropriately calibrate their trust in AI outputs. Specifically, research is needed for visualization
approaches aimed at surfacing doubt, encouraging verification, or preventing over-reliance. Because an over-trusting user is not motivated to initiate interaction with the model itself, we may need to shift our efforts from exploratory visualization to explanatory visualization, at least as a hook to incite skepticism.} 

\textcolor{Black}{Some methods from VXAI will continue to be valuable here, for example encodings to visualize class distributions, demographic representations, and feature correlations to reveal imbalances and biases in training datasets, or visual encodings to expose inconsistencies in feature importance. Other VXAI methods may have limited application, such as visual counterfactual analysis to reveal whether decision logic aligns with domain knowledge and causal reasoning, because of their reliance on interaction. Research is needed on how to make AI models more transparent in an obvious, understandable, and with an adequate level of friction. The additional benefit of showing this information without the need for interaction is that users may learn to expect this type of transparency in AI models.}


\subsection{Signal AI model failures} 
\textcolor{Black}{We have an obligation to clearly signal AI model failures and to support error pattern recognition. One idea would be showing failure case galleries to identify systematic mistakes potentially linked to model limitations or data quality issues. However, indicating the failure source is more complicated.} 

\textcolor{Black}{Furthermore, we could visualize decision surfaces and classification regions in new ways in order to expose whether models learn meaningful patterns or exploit spurious correlations and dataset artifacts. Research could also focus on visualizing unfavorable model responses to adversarial perturbations by default and investigate better, more compact visual encodings and designs for error analysis.}

\subsection{Signal current AI model limitations} 
\textcolor{Black}{
Third, we need to make current model limitations visible, preferably without requiring interaction. Visual comparisons of training, validation, and test performance, together with residual distributions, could highlight overfitting and performance disparities, revealing when models memorize rather than generalize and warning of potentially poor real-world performance.
Figure~\ref{fig:wentzel} illustrates the challenges in creating an explanatory visualization for a treatment decision recommendation, specifically designed to encourage the clinician to pause and reflect, without requiring interaction. Other design options considered here ranged from plain text warnings, similar to the ineffective legal disclaimers accompanying genAI overviews, to more salient markers, which obscured the prediction.} 

\textcolor{Black}{Further research is needed to communicate model uncertainty visually and help prevent over-reliance on uncertain predictions. Ideally, these visualizations would integrate with existing visualizations in an unobtrusive yet clear fashion. As above, making these limitations visible without disrupting users' workflows can encourage critical reflection rather than blind acceptance of model outputs.}

\subsection{Signal evolving limitations} 
\textcolor{Black} {Research is needed to help us detect model degradation as data distributions may change after deployment. Compact temporal visualizations could track changes in model predictions and performance metrics, while new visual designs could help reveal concept drift and its potential sources. As with the preceding agenda points, these visualizations should remain accessible to audiences with limited visualization literacy.}

    \subsection{Do this elegantly and unobtrusively}  \textcolor{Black} {Last but not least, the visual intervention points above need to be implemented unobtrusively, because clients clearly want to use the AI models for their functionality, and not get bogged down in caution messaging. At the same time, we still want to draw the viewer's attention, in order to enact slow thinking.}
    
    \textcolor{Black} {More generally, we would visually support colleague-level collaboration with the AI, where the AI model is presented as a knowledgeable but not infallible colleague. This is a tall order, plagued by prior human-computer-interaction failures, such as the notoriously beloved yet infamous Clippy Microsoft Office Assistant.}



\section{Conclusion}
We believe data visualization research can and should help support both trust and healthy skepticism in AI models. In this article, we have identified the rising phenomenon of over-reliance on AI models. We have tried to capture the benefits and risks associated with AI models. We made a case for being cautious and supporting healthy skepticism even in the case of performant AI models, and we identified a preliminary research agenda for the data visualization field in this direction. As long as we understand the AI is not a human and we judge its abilities by machine standards, it can be a powerful aid. 


Last but not least, we are experiencing an AI bubble through high AI valuations. Even so, the massive investment in foundational infrastructure will likely outlast the financial hype, accelerating long-term technology adoption, in the same way the dot-com wave has led to the information age. Recognizing risks helps ensure that humans retain the final, critical judgment, and are treating AI as a teammate rather than an autonomous superior. Regardless of the current AI wave evolution, supporting healthy skepticism in AI models is a valuable data visualization research direction.

\section*{Recommended readings (Sidebar 2)}
\noindent\hl{\textbf{Cancer Center Breaks With IBM Watson: }In 2017, the M. D. Anderson Cancer Center let its contract with IBM expire before anyone used the AI program Watson on actual patients:\\
R6. Schmidt, C., ``M. D. Anderson Breaks With IBM Watson, Raising Questions About Artificial Intelligence in Oncology.'' (2017).}\vspace{.5em}

\noindent\hl{\textbf{Healthy skepticism in AI models: }
Wentzel et al. (2024) identified the range of trust attitudes in AI models in the clinic, from critical to nearly blind acceptance, and introduced the concept of healthy skepticism:\\
R7. Wentzel, A., et al. ``DITTO: A visual digital twin for interventions and temporal treatment outcomes in head and neck cancer.'' (2024).}\vspace{.5em}

\noindent\hl{\textbf{De-skilling in endoscopists: }
This study (2025) documented the loss of skills in endoscopists after exposure to  colonoscopy AI:\\
R8. Budzyń, K., et al. ``Endoscopist deskilling risk after exposure to artificial intelligence in colonoscopy: a multicentre, observational study.'' (2025).}\vspace{.5em}

\noindent\hl{\textbf{Efficiency drop: }
This study (2025) showed that software developers took 19\% longer to complete tasks, despite believing AI makes them 20–24\% faster:\\
R9. Becker, J. et al. ``Measuring the impact of early-2025 AI on experienced open-source developer productivity.'' (2025).}
\vspace{.5em}

\noindent\hl{\textbf{Skill reduction in novice software engineers: }
This Anthropic study (2026) showed that novices gain speed at the cost of skill development:\\
R10. Shen, J. and Tamkin, A. ``How AI impacts skill formation.''  (2026).}\vspace{.5em}



\section{ACKNOWLEDGMENT}
We are grateful to the organizers and staff of the Dagstuhl seminar 26101 ``\textit{Contextualising Complexity -- Faithful Visualisations for Biology}'' for their support during the discussions. 
GEM's work is supported by U.S. NIH NCI R01CA258827, NIH UG3 TR004501, NIH NIDCR R01DE037101, NIH NIBIB R21EB037290, NIH NCI R01CA304878, U.S. NSF CNS-2320261, the American Cancer Society, and the UIC Institute for Health Data Science Research. 
MK was supported by the Deutsche Forschungsgemeinschaft (DFG, German Research Foundation) -- Project number 568853044.

\bibliographystyle{plain}
\bibliography{literature}

\begin{IEEEbiography}{Liz Marai} is a professor at University of Illinois Chicago. Contact her at gmarai@uic.edu.
\end{IEEEbiography}
\begin{IEEEbiography}{Marc Baaden} is a professor at CNRS France. 
\end{IEEEbiography}
\begin{IEEEbiography}{Michael Behrisch} is a professor at Utrecht University.
\end{IEEEbiography}
\begin{IEEEbiography}{Michael Krone} is a professor at Stuttgart Technical University.
\end{IEEEbiography}
\begin{IEEEbiography}{Pere-Pau Vazquez} is a professor at Universitat Politecnica de Catalunya.
\end{IEEEbiography}





\end{document}